\documentclass[10pt]{iopart}
\usepackage{graphicx}% Include figure files
\usepackage{bm}% bold math
\usepackage{iopams} %IOP maths
\usepackage{datetime}

\begin{document}

\title{Towards low-dimensional hole systems in Be-doped GaAs nanowires}

\author{A~R~Ullah$^{1}$, J~G~Gluschke$^{1}$, P~Krogstrup$^{2}$, C~B~S{\o}rensen$^{2}$, J~Nyg{\aa}rd$^{2}$ and A~P~Micolich$^{1}$}
\address{$^{1}$ School of Physics, University of New South Wales, Sydney NSW 2052, Australia}
\address{$^{2}$ Center for Quantum Devices, Nano-Science Center, Niels Bohr Institute, University of Copenhagen, Universitetsparken 5, DK-2100 Copenhagen, Denmark}
\ead{adam.micolich@nanoelectronics.physics.unsw.edu.au}

\begin{abstract}
GaAs was central to the development of quantum devices but is rarely used for nanowire-based quantum devices with InAs, InSb and SiGe instead taking the leading role. p-type GaAs nanowires offer a path to studying strongly-confined 0D and 1D hole systems with strong spin-orbit effects, motivating our development of nanowire transistors featuring Be-doped p-type GaAs nanowires, AuBe alloy contacts and patterned local gate electrodes towards making nanowire-based quantum hole devices. We report on nanowire transistors with traditional substrate back-gates and EBL-defined metal/oxide top-gates produced using GaAs nanowires with three different Be-doping densities and various AuBe contact processing recipes. We show that contact annealing only brings small improvements for the moderately-doped devices under conditions of lower anneal temperature and short anneal time. We only obtain good transistor performance for moderate doping, with conduction freezing out at low temperature for lowly-doped nanowires and inability to reach a clear off-state under gating for the highly-doped nanowires. Our best devices give on-state conductivity $95$~nS, off-state conductivity $2$~pS, on-off ratio $\sim~10^{4}$, and sub-threshold slope $50$~mV/dec at $T~=~4$~K. Lastly, we made a device featuring a moderately-doped nanowire with annealed contacts and multiple top-gates. Top-gate sweeps show a plateau in the sub-threshold region that is reproducible in separate cool-downs and indicative of possible conductance quantization highlighting the potential for future quantum device studies in this material system.
\end{abstract}

\submitto{\NT}
\newline
\noindent Version:~\today~~\currenttime
\maketitle
\ioptwocol

\section{Introduction}
GaAs has played a vital role in the development of quantum nanoelectronic devices over the past thirty years. The modulation-doped AlGaAs/GaAs heterostructure supports a quasi-two-dimensional electron gas with remarkably high electron mobility $>10^{5}$~cm$^2$/Vs~\cite{StormerAPL81}. This made AlGaAs/GaAs the system of choice for quantum Hall effect research~\cite{StormerRMP99}; thereafter it was central to work on 1D transport in quantum point contacts and quantum wires~\cite{BeenakkerSSP91, MicolichJPCM11}, Coulomb blockade and quantum interference in quantum dots~\cite{KouwenhovenNATO97}, and emerging designs for spin-based quantum information devices~\cite{vanderWielRMP03, HansonRMP07}. The semiconductor nanowire is another major advance in nanoelectronics. Born from the development of catalyst-assisted semiconductor growth methods in the 1960s~\cite{WagnerAPL64}, nanowires are being developed for applications from electronics and photonics to energy harvesting and biosensing~\cite{DasguptaAdvMat14}.

Nanowire-based quantum devices have primarily been developed using InAs/InP~\cite{BjorkNL04, FasthNL05, NadjPergeNat10, HeedtNL16}, InSb~\cite{NilssonNL09, MourikSci12, vanWeperenNL13} or SiGe nanowires~\cite{LuPNAS05, HuNatNano07, RoddaroPRL08}, despite GaAs being one of the first materials grown in nanowire form~\cite{HirumaJAP95}. Surface-states and their influence on carrier accumulation and contact ohmicity are largely responsible. Surface-states in InAs pin the surface Fermi energy above the conduction band minimum~\cite{NoguchiPRL91}. This results in electron accumulation at the nanowire surface enabling effective ohmic contact with negligible Schottky barrier via patterned deposition of metal contacts. Similar behaviour occurs for InSb nanowires~\cite{NilssonNL09}. In SiGe nanowires, the valence band offset causes hole accumulation in the Ge core without intentional doping; the holes can be accessed via annealed Ni contacts~\cite{LuPNAS05}. In each case, normally-on transistors are readily achieved without intentional doping using relatively simple contact formulations.

Surface-states in GaAs pin the surface Fermi energy mid-gap giving metal-semiconductor interfaces with a substantial Schottky barrier, typically $\sim 0.5$~eV, and a lack of intrinsic carrier accumulation~\cite{WaldropAPL84}. Ohmic contacts to GaAs are normally achieved using annealed alloys, e.g., NiGeAu for n-type contacts and PdZn or AuBe for p-type contacts~\cite{BacaTSF97}. Contact formation involves two key actions of the thermal annealing process. The first is diffusion of dopant species, i.e., Ge, Zn or Be, into the semiconductor. This degenerately dopes the adjacent semiconductor, producing a sufficiently narrow depletion region where tunnelling provides contact ohmicity~\cite{BraslauJVST81}. The second is strong spiking/penetration of Au into the semiconductor. This can be detrimental for contacts to bulk GaAs~\cite{GyulaiJAP71, KuanJAP83} but vital in heterostructure devices where contact access to buried conduction layers is required~\cite{TaylorJAP94}. Functional annealed contacts can be extremely difficult to obtain and optimise in nanowires due to their tiny volume and high aspect ratio; in many instances buckling and breakage of the nanowire and contact interfaces occurs. Additionally, GaAs nanowires have to be doped during growth for sufficient carrier density to conduct current. Doping nanowires is also complicated because there are numerous routes for dopant incorporation~\cite{GutscheJAP09} and common III-V dopants are amphoteric depending on incorporation site~\cite{DufouleurNL10}. The incorporation site is in turn dependent on growth mechanism (axial versus radial), conditions and the exposed facets -- the nanowire geometry invariably means numerous facets are exposed during growth.

\begin{figure}
\centering
\includegraphics[width=0.9\columnwidth]{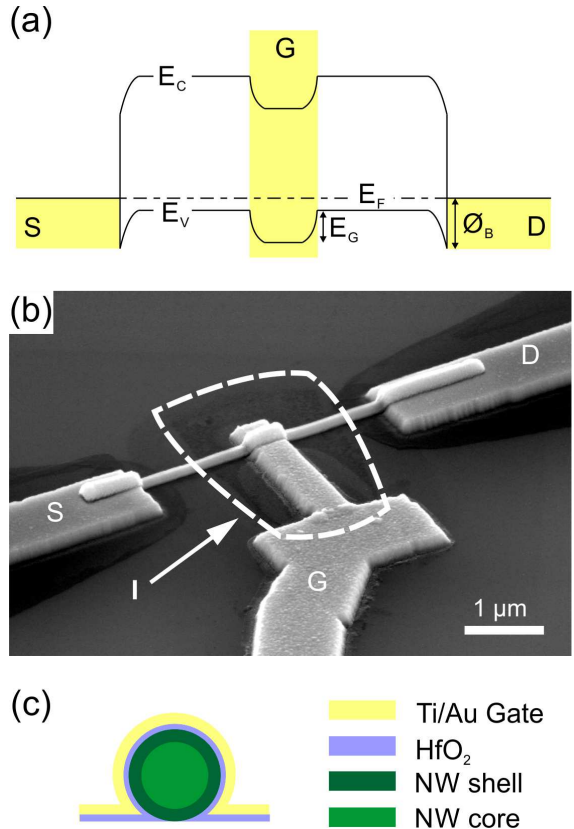}
\caption{(a) A simplified band diagram of the top-gate device at source-drain bias $V_{sd} = 0$~V. The source (S), drain (D), and top-gate (G) terminals are shown in yellow. The conduction band edge ($E_C$), valence band edge ($E_V$) and the Fermi energy ($E_F$) for a p-doped GaAs nanowire are shown. Surface states at the S/D terminals bend the bands, resulting in a Schottky barrier (${\phi}_{B}$). Finally, the effect of a positive top-gate voltage ($E_G$) is shown. (b) Scanning electron micrograph of a completed top-gate GaAs nanowire transistor. The S, D and G electrodes are all indicated along with the top-gate insulator (I). The n$^{+}$-Si substrate is used as a global back-gate for the device. (c) A cross-section of the top-gated part of the device.\label{fig:1}}
\end{figure}

GaAs nanowires are of significant interest for quantum device research~\cite{MorkotterNL15} despite these challenges to making functional high-quality devices. A particular area of opportunity is the study of strongly-confined 0D and 1D hole systems~\cite{CsontosPRB08, CsontosPRB09}. Holes in GaAs exhibit strong spin-orbit effects, behaving like spin-$\frac{3}{2}$ particles, by virtue of the $p$-like nature of GaAs valence band states~\cite{WinklerBook03}. This led to the observation of highly anisotropic Zeeman spin-splitting in 1D hole systems ~\cite{KlochanNJP09, ChenNJP10, SrinivasanNL13} and correspondingly anisotropic Kondo physics in hole quantum dots~\cite{KlochanPRL11} made using conventional planar p-type AlGaAs/GaAs heterostructures. A major difficulty in these devices is the ability to achieve strong confinement and access the few-electron limit~\cite{KlochanAPL10}. The unique geometry offered by GaAs nanowires may overcome these difficulties. In this paper we report on the development of field-effect transistor devices based on Be-doped p-type GaAs nanowires grown by Molecular Beam Epitaxy (MBE). Our devices feature both doped silicon back-gates~\cite{DuanNat01} and oxide-insulated top-gate structures~\cite{PfundAPL06}, the latter with strong potential for realising functional quantum devices. Figure~\ref{fig:1}a shows a band diagram of a p-doped GaAs top-gated transistor. We focus here mostly on electrical transistor performance relative to dopant concentration and contact preparation, but also present preliminary data indicative of possible 1D conductance plateaus in our nanowire devices.

\section{Fabrication and Measurement}
Our homostructure core-shell GaAs nanowires were grown by MBE on $(111)$ Si~\cite{CasadeiAPL13} using the self-catalysed Ga-assisted method~\cite{FontcubertaAPL08, ColomboPRB08}. The undoped core was grown at $630^{\circ}$C using an As$_{4}$ source and V/III flux ratio of $60$ or $30$ for $45$~minutes. A Be-doped shell was then grown at $465^{\circ}$C using As$_{2}$ and a V/III ratio of $150$ for $30$~minutes, giving final structures with diameter $120-200$~nm and length $6-10~\mu$m. We studied nanowires with three different shell doping densities $n_{Be}~=~1~\times~10^{18}$~cm$^{-3}$, $5~\times~10^{18}$~cm$^{-3}$ and $1.5~\times~10^{19}$~cm$^{-3}$, which we will refer to as low $n_{Be}$, moderate $n_{Be}$ and high $n_{Be}$, respectively. $n_{Be}$ was estimated using four-terminal measurements on grown nanowires~\cite{CasadeiAPL13}.

Device fabrication was performed on a degenerately-doped Si wafer capped with a bilayer oxide consisting of $100$~nm thermally-grown SiO$_{2}$ and $10$~nm HfO$_{2}$ deposited by Atomic Layer Deposition (ALD). The bilayer oxide is not essential and is an artefact of our other nanowire device projects~\cite{StormNL12}. The wafer was pre-patterned with alignment markers for Electron-Beam Lithography (EBL) and Ti/Au interconnects. Nanowires were randomly deposited by dry transfer and then located by darkfield optical microscopy to facilitate deposition of contacts and gating structures direct to a selected nanowire. In this study, two different types of devices were made. The simplest devices feature a pair of source and drain contacts made with $200$~nm of 99:1 Au:Be alloy (ACI alloys) defined by EBL and vacuum thermal evaporation. In some instances the AuBe contacts were annealed using an Ulvac-Riko Mila 4000 rapid thermal annealer with carbon susceptor at temperatures between $250^{\circ}$C and $350^{\circ}$C for times up to $120$~s. The anneal temperature was deliberately kept below $350^{\circ}$C to avoid the onset of strong Au penetration/spiking~\cite{GyulaiJAP71} from destroying the nanowire, although mild Au penetration into GaAs still occurs well below $350^{\circ}$C~\cite{YoshiieTSF84}. Gating in these devices was achieved using the degenerately doped Si substrate~\cite{DuanNat01}. Our more complex devices also feature a top-gate, which is produced in two steps following fabrication of the AuBe contacts. The first step was deposition of an EBL-defined $10$~nm HfO$_{2}$ top-gate insulator by ALD. The second step was deposition of an EBL-defined $20/180$~nm Ti/Au gate electrode. Figure~\ref{fig:1}b shows a scanning electron micrograph of a completed top-gate device with the HfO$_{2}$ insulator highlighted by the dashed line. More complete details of the fabrication process are provided in the Supplementary Information. Figure~\ref{fig:1}c shows a cross section through the top-gated section of the device.

Device measurements were performed using either a probe station (Signatone) in a Faraday cage in ambient conditions, a home-built dipstick loaded into a He dewar or an Attocube attoDry 2100 variable-temperature system with base temperature $T~=~1.5$~K. DC measurements were used to characterise the devices. In Section \ref{sec:contacts}, a source-drain bias $V_{sd}$ was applied using a Keithley K236 source-measure unit, which simultaneously measures the source current $I_{s}$. The drain current $I_{d}$ is measured using a Keithley K6517A electrometer. Unless mentioned explicitly, $I_{s}~=~I_{d}$, which we refer to as the source-drain current $I_{sd}$ hereafter. In Section \ref{sec:topGate}, a Yokogawa GS200 was used to set $V_{sd}$ and a Keithley K6517A was used to measure $I_{sd}$. The back-gate voltage $V_{bg}$ and top-gate voltage $V_{tg}$ were applied using Keithley K2401 source-measure units to enable continuous monitoring of gate leakage current $I_{g}$.

\section{AuBe as a contact metal for Be-doped nanowires} \label{sec:contacts}
The AuBe ohmic contacts are one of the key contributors to the stability and performance of the devices in Refs.~\cite{KlochanNJP09, ChenNJP10, SrinivasanNL13, KlochanPRL11}. These contacts give reliable linear $I-V$ with relatively high yield in these devices subject to appropriate optimization of anneal conditions~\cite{ClarkeJAP06}. This made AuBe an interesting contact metal to try with Be-doped GaAs NWs, with the intention that the contacts could be annealed to supplement the Be density local to the contact and facilitate ohmic p-type contact to GaAs nanowires as the shell doping density is reduced. Thus we now focus on our simple back-gated samples to look at contact efficacy versus $n_{Be}$ and anneal conditions. We begin by looking at unannealed AuBe contacts and then move on to discuss how annealing alters contact performance.

\begin{figure}
\centering
\includegraphics[width=0.9\columnwidth]{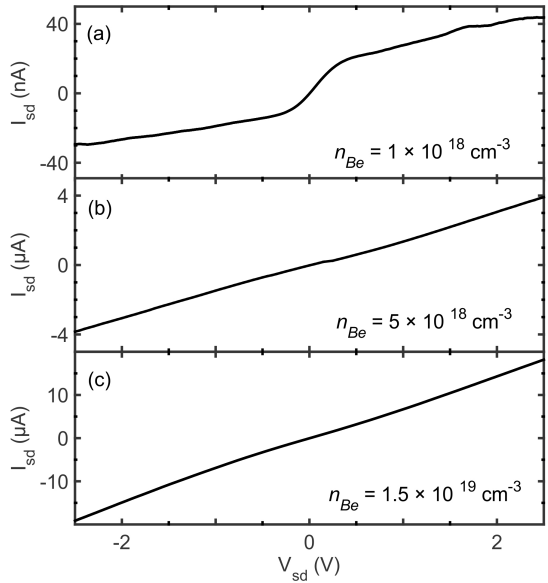}
\caption{Plot of source-drain current $I_{sd}$ vs source-drain bias $V_{sd}$ at $V_{bg} = 0$~V for doping densities (a) $n_{Be} = 1~\times~10^{18}$~cm$^{-3}$, (b) $n_{Be} = 5~\times~10^{18}$~cm$^{-3}$, and (c) $n_{Be} = 1.5~\times~10^{19}$~cm$^{-3}$. The contacts are unannealed in all three cases. \label{fig:2}}
\end{figure}

Figure~\ref{fig:2} shows typical $I_{sd}$ versus $V_{sd}$ characteristics for the three different nanowire doping densities investigated with unannealed AuBe contacts. The main feature indicating good contact performance is that $I_{sd}$ increases linearly rather than exponentially as $|V_{sd}|$ is increased from zero -- the former is characteristic of ohmic contacts whereas the latter is characteristic of diode-like Schottky contacts. The $I_{d}$ versus $V_{sd}$ characteristic for low $n_{Be}$ in Fig.~2(a) is non-linear but with the opposite curvature to Schottky contact behaviour -- the curvature here arises due to saturation and is normal for field-effect transistors (FETs). This form is observed in $85\%$ of our low $n_{Be}$ devices, with around $5\%$ showing rectifying contacts instead. The comparative linearity for Figs.~2(b/c) where $n_{Be}$ is much higher is expected because saturation voltage is proportional to doping density in FETs~\cite{SzeBook06}, pushing saturation outside the $\pm 2.5$~V measurement range. The increased $n_{Be}$ also reduces the Schottky barrier depletion width~\cite{CasadeiAPL13} enabling tunneling to contribute strongly to the contact conduction process~\cite{BraslauJVST81}. Correspondingly, the conductance $G$ increases from $48~\pm~23$~nS for low $n_{Be}$ to $1.5~\pm~0.9~\mu$S for moderate $n_{Be}$ and $9.4~\pm~2.9~\mu$S at high $n_{Be}$, respectively. We also examined the symmetry of the $I_{d}$ versus $V_{sd}$ characteristics as an additional indicator of contact quality. We defined an $I_{d}$ versus $V_{sd}$ characteristic as `symmetric' if both: a) the $I_{sd}$ at $V_{sd} = \pm 2.5$~V, and b) the integrals $\int I_{sd}dV_{sd}$ for $-2.5 < V_{sd} < 0$~V and $0 < V_{sd} < 2.5$~V differ by less than $25\%$. Across the $30-35$ devices measured at each $n_{Be}$, we found $17\%$ gave symmetric $I_{d}$ versus $V_{sd}$ characteristics at low $n_{Be}$, $44\%$ at moderate $n_{Be}$, and $100\%$ at high $n_{Be}$. We conclude that the high $n_{Be}$ nanowires give the best contact performance.

\begin{figure}
\centering
\includegraphics[width=0.9\columnwidth]{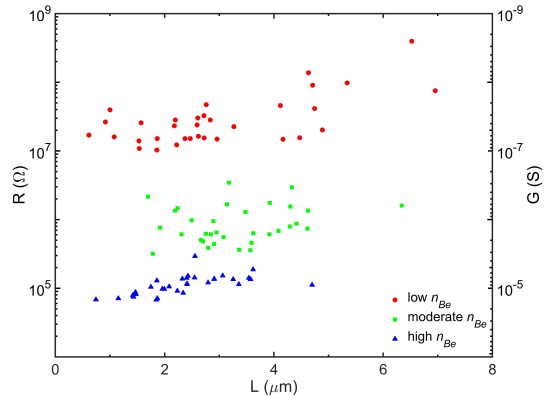}
\caption{Plot of two-terminal resistance $R$ vs channel length $L$ for $n_{Be} = 1~\times~10^{18}$~cm$^{-3}$ (red circles), $n_{Be} = 5~\times~10^{18}$~cm$^{-3}$ (green squares) and $n_{Be} = 1.5~\times~10^{19}$~cm$^{-3}$ (blue triangles).\label{fig:3}}
\end{figure}

We measured a sufficient number of devices with unannealed contacts that we can also assess resistance $R$ versus channel length $L$ for the three different $n_{Be}$; the data is plotted in Fig.~\ref{fig:3}. The resistance shows no clear trend with length for low and moderate $n_{Be}$, as expected if the contact resistance dominates the channel conductivity. We estimate the contact resistance for the low and moderate $n_{Be}$ devices to be $10$~M$\Omega$ and $200$~k$\Omega$, respectively, by extrapolating to $L = 0 \mu$m. There is a clear relationship between $R$ and $L$ for high $n_{Be}$, as expected when the contact resistance is no longer dominant. We estimate the contact resistance for high $n_{Be}$ to be $29~\pm~15$~k$\Omega$.

\begin{figure}
\centering
\includegraphics[width=0.9\columnwidth]{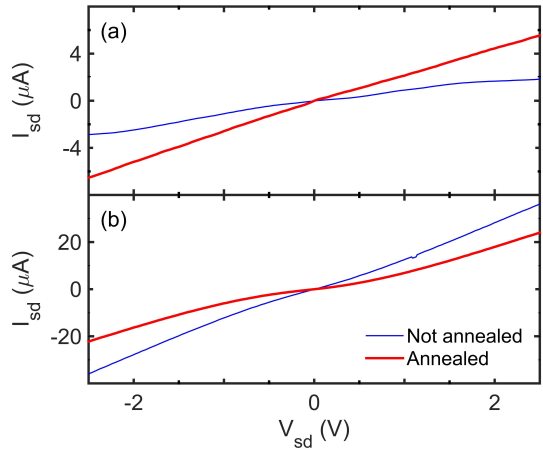}
\caption{Plots of $I_{sd}$ vs $V_{sd}$ at $V_{bg} = 0$~V for AuBe contacts that have (a) improved on a moderate $n_{Be}$ device and (b) deteriorated on a high $n_{Be}$ device after annealing. In both cases, data is shown immediately prior to annealing (thin blue line) and immediately after annealing (thick red line). \label{fig:4}}
\end{figure}

We now turn our attention to annealed AuBe contacts. The typical anneal for AuBe contacts to planar p-type AlGaAs/GaAs heterostructures is $450-550^{\circ}$C for $60-120$~s~\cite{ClarkeJAP06}. However, in that instance, strong diffusion of ohmic spikes to a significant depth $\sim 100$~nm is required to access the 2D hole layer. Such an anneal is too aggressive for our nanowires. We instead focus in the $250-350^{\circ}$C range where Au penetration is not so aggressive~\cite{GyulaiJAP71} for times of $30-120$~s to obtain Be diffusion without severely detrimental effects, e.g., buckling and nanowire breakage. These anneal conditions are not too dissimilar to those used for NiGeAu contacts to n-type GaAs/AlGaAs core-shell nanowires by Mork\"{o}tter {\it et al.}~\cite{MorkotterNL15}. For each set of anneal conditions, we measure the source-drain characteristics immediately prior to annealing and immediately thereafter, with $V_{bg} = 0$~V. Annealing generally produces one of two effects -- either the contact is improved, becoming more linear with higher $I_{sd}$ at fixed $V_{sd}$, or it is deteriorated, giving lower $I_{sd}$ at fixed $V_{sd}$ and/or reduced symmetry. Figure~\ref{fig:4}(a) and (b) show typical examples of improved and deteriorated contacts, respectively.

\begin{figure*}
\centering
\includegraphics[width=0.9\textwidth]{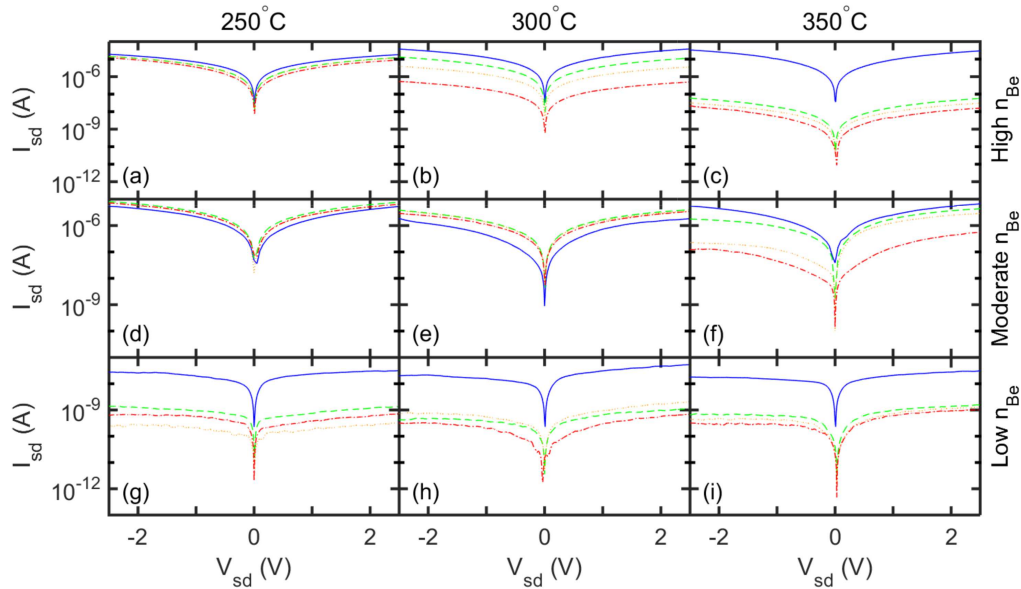}
\caption{Plots of $I_{sd}$ vs $V_{sd}$ at $V_{bg} = 0$~V for (a) $T = 250^{\circ}$C, high $n_{Be}$, (b) $T = 300^{\circ}$C, high $n_{Be}$, (c) $T = 350^{\circ}$C, high $n_{Be}$, (d) $T = 250^{\circ}$C, moderate $n_{Be}$, (e) $T = 300^{\circ}$C, moderate $n_{Be}$, (f) $T = 350^{\circ}$C, moderate $n_{Be}$, (g) $T = 250^{\circ}$C, low $n_{Be}$, (h) $T = 300^{\circ}$C, low $n_{Be}$, (i) $T = 350^{\circ}$C, low $n_{Be}$. In each panel, there are four traces presented: unannealed (blue solid), annealed for $30$~s (green dashed), annealed for $60$~s (orange dotted) and annealed for $120$~s (red dot-dashed). Corresponding linear plots are shown in Supplementary Figure~1.~\label{fig:5}}
\end{figure*}

Figure~\ref{fig:5} shows the evolution of $I_{sd}$ versus $V_{sd}$ with annealing time for three different anneal temperatures at each of the three doping densities. The logarithmic scale for $I_{sd}$ is used to better highlight changes in $I_{sd}$ with asymmetry evident as different saturation levels at high positive/negative $V_{sd}$. The corresponding data plotted on a linear scale is presented in the Supplementary Information. Before elaborating on specifics, our overall finding is that while improvement in contact quality can be obtained in some circumstances, annealing generally proves detrimental. The best outcome occurs for moderate $n_{Be}$ at $T = 300^{\circ}$C (Fig.~\ref{fig:5}(e)) where significant improvement can be achieved for minimal anneal time. A similar outcome is attained for moderate $n_{Be}$ at $T~=~250^{\circ}$C (Fig.~\ref{fig:5}(f)). Otherwise annealing generally provides little improvement, e.g., high $n_{Be}$ at $T~=~250^{\circ}$C (Fig.~\ref{fig:5}(g)), or makes the contacts drastically worse (all other panels). The anneal is uniformly detrimental for low $n_{Be}$, which unfortunately is where improved contacts are most needed. Given this result, we performed a more extended study of anneal efficacy at moderate $n_{Be}$. Table~\ref{tab:1} presents the yield of improved contacts versus anneal temperature and time, with the optimum outcome being an $83\%$ yield for annealing at $250^{\circ}$C for $60$~s.

\begin{table}
    \centering
    \begin{tabular}{c c c c c c}
    \hline
    $T(^{\circ}$C) & \multicolumn{4}{c}{Annealing time} \\
    & & $30$~s & $60$~s & $90$~s & $120$~s \\
    \hline
    250 & & 75\% & 83\% &  & 75\% \\
    300 & & 73\% & 45\% & 36\% & 36\% \\
    350 & & 18\% & 18\% & 10\% & 10\% \\
    \hline
    \end{tabular}
    \caption{This table shows the yield of improved contacts by annealing for the $n_{Be} = 5 \times 10^{18}$~cm$^{-3}$ (moderately doped) nanowires.~\label{tab:1}}
\end{table}

We can explain our data in Fig.~\ref{fig:5} on the basis that the diffusion of grown-in Be dopants within the nanowire is much faster than the diffusion of Be from the AuBe contacts into the nanowire. We are unable to find data for Be out-diffusion rate from AuBe alloy on GaAs but Be is well-known for its high diffusion rate as a dopant species in GaAs~\cite{ManfraAPL05, MasuJAP83, IlegemsJAP77}. Under this scenario, for the low $n_{Be}$ samples, loss of Be from shell to core should both deteriorate the contacts, due to a rise in surface depletion width~\cite{CasadeiAPL13}, and reduce the overall nanowire conductivity. For moderate and high $n_{Be}$, small amounts of dopant diffusion consistent with short anneals at relatively low temperature should not significantly deteriorate the contacts, and might well improve them slightly, as observed in Fig.~\ref{fig:5}(d,e). That said, higher temperatures and longer anneal times should tip the balance toward worse contact performance, as in the low $n_{Be}$ case, due to net diffusion of Be out of the shell.

\section{Top-gated Be-doped GaAs nanowire transistors}\label{sec:topGate}
We now turn our attention to obtaining local gating for our Be-doped GaAs nanowires. Local gating requires patterned gate electrodes, which are most readily obtained by depositing an oxide layer by ALD to insulate the nanowire followed by EBL to define metal top-gate electrodes. The oxide can either be a conformal coating applied to the nanowires whilst standing vertically on the growth substrate, or a patterned oxide over the partially-complete device structure. The InAs native oxide has been used to insulate local gate electrodes previously~\cite{PfundAPL06} but these gates only perform well, i.e., without strong current leakage, for small gate biases. Ultimately, the need for a gate oxide in In-based nanowires is driven by the small Schottky barrier arising from surface-state pinning~\cite{NoguchiPRL91}. In contrast the oxide should, in principle, be optional for GaAs nanowires because the mid-gap surface pinning induces a stronger Schottky barrier. Although this is true for AlGaAs/GaAs heterostructure devices, it does not work so well for our GaAs nanowire transistors, presumably due to the shell doping. After fabricating such a device with low $n_{Be}$ nanowires, we found that the gate leakage current $I_{g}~\sim~400$~pA at the peak $I_{sd}$ of 1~nA. In contrast, our gate leakage current is always below $10$~pA for top-gates featuring a HfO$_{2}$ insulator.

\begin{figure}
\centering
\includegraphics[width=0.9\columnwidth]{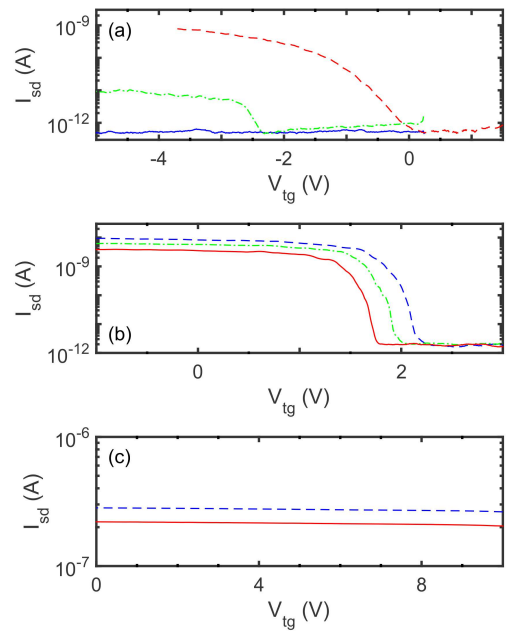}
\caption{Plots of $I_{sd}$ vs $V_{tg}$ for (a) low $n_{Be}$ at $T~=~250$~K (dashed red), $T~=~218$~K (dot-dashed green) and $T~=~155$~K (solid blue); (b) moderate $n_{Be}$ at $T~=~4$~K for back-gate voltages $V_{bg}~=~-3$~V (dashed blue), $V_{bg}~=~0$~V (dot-dashed green) and $V_{bg}~=~+3$~V (solid red); and (c) high $n_{Be}$ at $T~=~4$~K for back-gate voltages $V_{bg} ~=~0$~V (dashed blue), $V_{bg}~=~+20$~V (solid red). $V_{sd}~=~100$~mV for all traces.~\label{fig:6}}
\end{figure}

Since our objective is quantum transport studies, we made top-gate transistors for each $n_{Be}$ and examined their performance as the temperature $T$ was reduced to $4$~K. The aim was to find the lowest $n_{Be}$ giving the following desirable properties at low $T$: 1. reasonably symmetric and linear $I_{sd}$ versus $V_{sd}$ characteristics; 2. on-state conductance of order $100$~nS or more; and 3. off-state conductance below $10$~nS (i.e., on-off ratio better than $10^3$). The desire for the lowest $n_{Be}$ possible is to minimise dopant scattering and thereby maximise ballistic transport length.

\begin{figure}
\centering
\includegraphics[width=0.9\columnwidth]{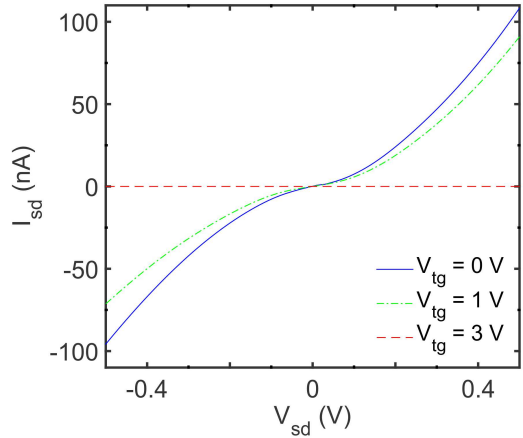}
\caption{Plots of $I_{sd}$ vs $V_{sd}$ at $V_{tg}~=~0$~V (blue solid line), $V_{tg}~=~+1$~V (green dot-dashed line) and $V_{tg}~=~0$~V (red dashed line) for moderate $n_{Be}$, $T~=~4$K and $V_{bg} = 0$~V.~\label{fig:7}}
\end{figure}

Figure~\ref{fig:6} shows $I_{sd}$ versus $V_{tg}$ for devices with different $n_{Be}$. At low $n_{Be}$, the room temperature on-state conductivity $G~\sim~8$~nS is low, and as we reduce $T$ conduction freezes out (Fig.~\ref{fig:6}(a)). By $T~=~155$~K we no longer see distinct on- and off-states. Thus, combined with the poor $I_{sd}$-$V_{sd}$ characteristics shown earlier, these nanowires are unsuitable for quantum transport studies. The high $n_{Be}$ nanowires are also unsuitable. Figure~\ref{fig:6}(c) shows data for a high $n_{Be}$ device at $T~=~4$~K. While the on-state conductivity is high, and the $I_{sd}$ versus $V_{sd}$ characteristics are good, the on-off ratio is poor at $\sim1.1$ with a high off-state current, i.e., we cannot properly deplete the high $n_{Be}$ nanowires with a metal top-gate. In contrast, the moderate $n_{Be}$ devices show considerable promise for low $T$ quantum transport applications. Figure~\ref{fig:6}(b) shows data obtained at $T~=~4$~K for three different back-gate voltages $V_{bg}$ in a device annealed at $T~=~250^{\circ}$C for $60$~s. These devices show good on-state and off-state conductivity, $95$~ns and $2$~pS respectively, and high on-off ratio $\sim~10^{4}$. The top-gate sub-threshold slope is $50$~mV/decade at $V_{bg}~=~0$~V at $T~=~4$~K. We found the top-gate subthreshold slope to be an order of magnitude better than is typical using the back-gate. We also found scope for tuning the top-gate threshold voltage using the back-gate (Fig.~6(b)). We obtain a typical field-effect mobility of only $\sim2$~cm$^{2}$/Vs at $T~=~4$~K but note that this will be severely contact-limited for our devices and is unlikely to be a reliable measure of the true channel performance. The moderate $n_{Be}$ devices also give decent contact performance at $T~=~4$~K. As Fig.~\ref{fig:7} shows, $I_{sd}$ versus $V_{sd}$ is somewhat non-linear but at least symmetric, and with sufficient conductance at reasonable $V_{sd}$ to viably use for further studies.

\begin{figure}
\centering
\includegraphics[width=0.9\columnwidth]{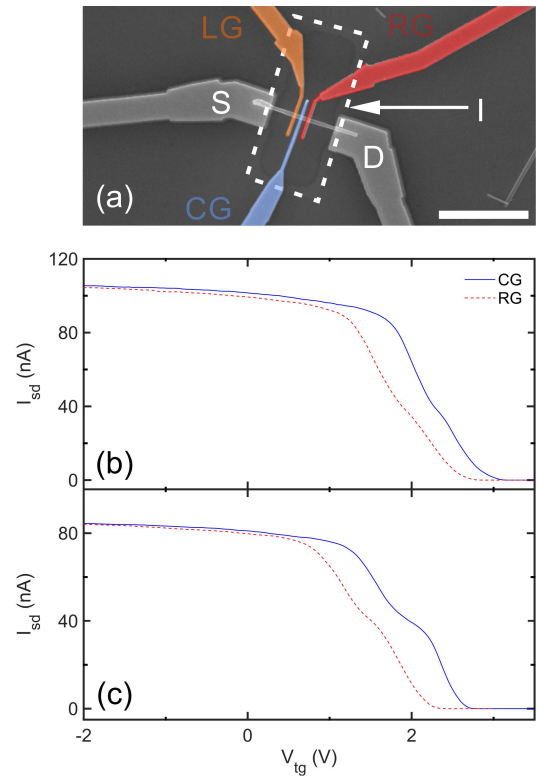}
\caption{(a) Scanning electron micrograph of a three-gate nanowire transistor made using a moderate $n_{Be}$ GaAs nanowire and AuBe contacts annealed at $300^{\circ}$C for $30$~s. The source (S), drain (D), insulator (I), left-gate (LG) - shaded orange, centre-gate (CG) - shaded blue, and right-gate (RG) - shaded red, are indicated. The three gate electrodes are $200$~nm wide with $200-300$~nm spacing and insulated from the nanowire by $10$~nm of HfO$_2$. The scale bar represents $5~\mu$m. (b) $I_{sd}$ vs $V_{tg}$ for CG (solid blue) and RG (dashed red) for the device in (a). Measurements were obtained at $T~=~1.5$~K with $V_{sd}~=~500$~mV (rms), $V_{bg}~=~0$~V and no magnetic field applied. (c) $I_{sd}$ vs $V_{tg}$ for CG (solid blue) and RG (dashed red) obtained on a separate cooldown. The dashed red traces in both (b) and (c) are horizontally offset by $-0.5$~V for clarity. Any top-gate not being actively swept is held at $V_{tg}~=~-2$~V (strong on-state) to prevent them adversely influencing the measurements.~\label{fig:8}}
\end{figure}

We made the device shown in Fig.~\ref{fig:8}(a) to demonstrate the potential for quantum transport studies using the moderate $n_{Be}$ nanowires with contacts annealed at $T = 300$~$^{\circ}$C for 30~s. The device architecture features three narrow top-gates, each designed to define a short constriction in the nanowire, and together form a tunable quantum dot. Figure~\ref{fig:8}(b) shows $I_{sd}$ versus $V_{tg}$ obtained at $T~=~1.5$~K for the two gates RG and CG -- unfortunately the third gate LG functions poorly (see Supplementary Fig.~S4) and was not considered further for this work. The characteristics for RG and CG show clear on- and off-states, strong sub-threshold behaviour, and curiously, a clear plateau at a common intermediate conductance for both gates. Figure~\ref{fig:8}(c) shows the same measurement obtained on a separate cooldown. The reduced on-state $I_{sd}$ is due to aging of the sample between the two cooldowns, which were separated in time by several weeks. The plateau structure appears similarly in both cooldowns. This suggests the plateau's origin is not a quantum interference resonance arising from the ionized doping potential, since this should reconfigure~\cite{SeePRL12} upon warming to room temperature given Be is a shallow acceptor in GaAs.~\cite{LewisPRB96}

\begin{figure}
\centering
\includegraphics[width=0.9\columnwidth]{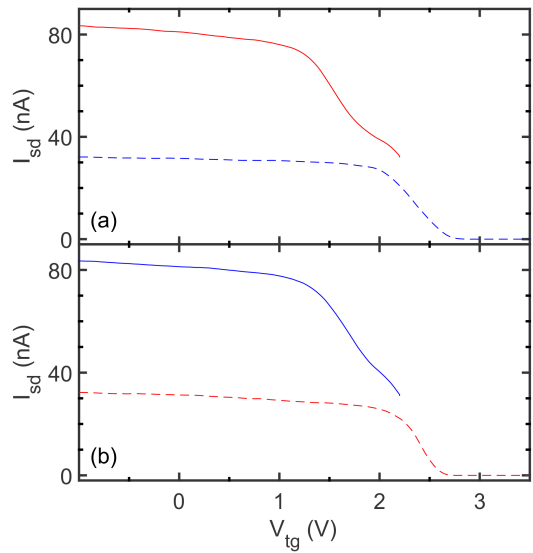}
\caption{Plots of $I_{sd}$ vs $V_{tg}$ for (a) RG taken first to just below first plateau $V_{RG}~=~+2.2$~V with $V_{CG}~=~-2$~V (solid red) followed by a sweep of CG to pinch-off with $V_{RG}~=~+2.2$~V (dashed blue), and (b) CG taken first to just below first plateau $V_{CG}~=~+2.2$~V with $V_{RG}~=~-2$~V (solid blue) followed by a sweep of RG with $V_{CG}~=~+2.2$~V (dashed red). Measurements were obtained at $T~=~300$~mK with $V_{sd}~=~500$~mV (rms), $V_{bg}~=~0$~V and no magnetic field applied.~\label{fig:9}}
\end{figure}

Figure~9 shows data obtained to further investigate the plateau structures in Fig.~8. The data in Fig.~9(a) is obtained in two stages. We start with CG and RG in a strong on-state by setting $V_{CG}~=~V_{RG}~=~-2$~V. First we sweep RG to just below the plateau, i.e., $V_{RG}~=~+2.2$~V. Assuming the plateau indicates quantised conductance, this would mean that just a single 1D mode is transmitted through the region under RG. Second, we sweep CG from $V_{CG}~=~-2$~V to pinch-off ($I_{sd}~=~0$). Interestingly, the plateau in the CG trace disappears in this case (c.f. data in Fig.~8(c)). In Fig.~9(b) we repeat this measurement with the gates reversed -- sweep CG to just below the plateau, then sweep RG to pinch-off. Here also the plateau disappears for the RG trace. This behaviour is consistent with that observed in a pair of QPCs separated by less than the ballistic length~\cite{WharamJPC88}, where conservation of 1D sub-band index in transport means the series resistance is determined by whichever QPC passes the fewest 1D sub-bands. This has also recently been observed in InAs nanowire devices~\cite{HeedtNL16}. In the context of Fig.~9(a), for example, it means that once we drive RG to below the first plateau, then only a single 1D subband can pass under CG. This single 1D subband remaining in transport at $V_{CG}~=~+2.2$~V is pinched-off as a single step as $V_{RG}$ is driven positive. The same holds on reversing the gates. The behaviour we observe here lends further support for the plateau we observe being caused by 1D conductance quantisation.

Future work will focus more closely on quantum transport effects in these p-GaAs nanowire devices. This will include studies of larger diameter nanowires seeking to obtain more than the single plateau we observe in the structures reported here as well as Coulomb-blockaded quantum dot devices for studying spin-orbit effects in the single-hole limit.

\section{Conclusion}
We reported on the development of nanowire transistors featuring Be-doped p-type GaAs nanowires and AuBe alloy contacts towards making nanowire-based hole quantum devices. Devices with traditional doped-substrate back-gates and EBL-defined metal/oxide top-gates were produced for three different Be-doping densities $n_{Be}$ and a range of AuBe contact processing recipes. A key focus was the comparison of unannealed contacts with those annealed at $250-350^{\circ}$C for up to $120$~s using a rapid thermal annealer. Annealing only brings small improvements for the moderately-doped devices under conditions of lower anneal temperature and short anneal time. Otherwise, annealing generally proves detrimental. We attribute this to the diffusion of Be dopants from the shell being much faster than the out-diffusion of Be from the AuBe alloy into the GaAs. The performance of the top-gate transistors is also $n_{Be}$-dependent. Conduction in the lowest $n_{Be}~=~1~\times~10^{18}$~cm$^{-3}$ nanowires freezes out below $\sim~100$~K making these devices unsuitable for quantum device applications. The highest $n_{Be}~=~1.5~\times~10^{19}$~cm$^{-3}$ devices cannot be fully depleted by our top-gates and give poor on-off ratio $\sim 1.1$ making them also unsuitable. The moderate $n_{Be}~=~5~\times~10^{18}$~cm$^{-3}$ devices with contacts annealed $T = 250^{\circ}$C for 60~s give good on-state conductivity $95$~nS, off-state conductivity $2$~pS, on-off ratio $\sim~10^{4}$, and sub-threshold slope $50$~mV/dec at $T~=~4$~K, making them well suited for quantum device studies. Lastly, we made a device featuring a moderate $n_{Be}$ nanowire with annealed contacts and multiple top-gates. Top-gate sweeps show a plateau in the sub-threshold region indicative of possible conductance quantization that is reproducible in separate cool-downs, demonstrating the potential of our device structures for future quantum transport studies.

\ack This work was funded by the Australian Research Council (ARC), the University of New South Wales, Danish National Research Foundation and the Innovation Fund Denmark. APM acknowledges an ARC Future Fellowship (FT0990285) and an ARC Discovery Grant (DP110103802). This work was performed in part using the NSW node of the Australian National Fabrication Facility (ANFF).

\end{document}